\documentclass[notitlepage,nofootinbib]{revtex4-1}

\usepackage{amsmath}
\usepackage{amsfonts,amssymb}
\usepackage{hyperref}
\usepackage{graphicx}
\usepackage{natbib}
\usepackage[utf8]{inputenc}
\usepackage{xcolor}
\usepackage{bm}
\usepackage{multirow}
\usepackage{diagbox}
\usepackage{hhline}
\usepackage[normalem]{ulem}

\usepackage[normalem]{ulem}

\usepackage{xcolor}

\begin{document}

\title{Induced non-local cosmology}

\author{Leonardo Giani$^a$}
\email{l.giani@uq.edu.au}
\affiliation{$^a$ School of Mathematics and Physics, The University of Queensland, Brisbane, QLD 4072, Australia}

\author{Oliver F. Piattella$^{b,c}$}
\email{of.piattella@uninsubria.it}

\affiliation{$^b$ Dipartimento di Scienza e Alta Tecnologia, Universit\`a degli Studi dell'Insubria e INFN, via Valleggio 11, I-22100 Como, Italy\\
$^c$ N\'ucleo Cosmo-ufes, Universidade Federal do Esp\'irito Santo, avenida F. Ferrari 514, 29075-910 Vit\'oria, Esp\'irito Santo, Brazil}

\begin{abstract}
We investigate the cosmological implications of an effective gravitational action, inspired by Sakharov's idea of induced gravity, containing non-local contributions from the operator $\left(\Box +\beta \right)^{-1} R$. The $\beta$ term is a novel feature in the panorama of non-local models of gravity, and arises naturally within Sakharov theory from the potential of a non-minimally coupled scalar field after a spontaneous symmetry breaking takes place. In this class of models the non-local contribution can acquire an oscillatory behaviour, thereby avoiding the divergence of the Hubble parameter at late times, which is another characteristic feature of the non-local models treated in the literature. Furthermore, the effective gravitational coupling $G_{\rm eff}$ inherits the oscillatory behaviour, resulting in alternating epochs of stronger and weaker gravity.
This framework is argued to have potentially interesting implications for the $H_0$ and the $\sigma_8$ tensions.  
\end{abstract}

\maketitle

\section{Introduction}

Non-local gravity has received interest because it is able to provide a natural explanation for the observed accelerated expansion of the universe without introducing a new mass scale, see e.g. Refs. \cite{Wetterich:1997bz,Woodard:2014iga, Maggiore:2016gpx, Belgacem:2017cqo, Giani:2019xjf, Giani:2019vjf, Capozziello:2021krv} for reviews. Amongst the simplest non-local corrections we can add to the Einstein-Hilbert action there are those containing the inverse of some differential operator acting on the Ricci scalar, such as $\Box^{-1}R$ \cite{Vardanyan:2017kal}, $R$ times a function (dubbed distortion function) of $\Box^{-1}R$ in the Deser-Woodard (DW) model \cite{Deser:2007jk, Amendola:2019fhc}, or even powers of the latter like in the $R\;\Box^{-2}R$ model \cite{Maggiore:2014sia,Amendola:2017qge}. Besides explaining dark energy, non-local models have been studied also in different frameworks such as the propagation of gravitational waves \cite{Belgacem:2019lwx, Capozziello:2021bki}, bouncing cosmology \cite{Jackson:2021mgw,Chen:2019wlu}, black holes \cite{Kumar:2018pkb,Fu:2022yrs} and as solution of the cosmological constant problem \cite{Nojiri:2010pw,Zhang:2011uv}.

The effectiveness of such non-local deformations in providing an accelerated expansion is readily explained by looking at their (retarded) integral representations on a Friedmann-Lemaître-Robertson-Walker (FLRW) background. Consider, for example, the simplest non-trivial DW term \cite{Deser:2007jk}:
\begin{equation}\label{inverseboxR}
    \frac{1}{\Box}R(t) = -\int_0^tdt'\frac{1}{a^3(t')}\int_0^{t'}dt'' a^3(t'')R(t'')\;, 
\end{equation}
where $a$ is the scale factor. If radiation dominates the energy density of the Universe, i.e. assuming that the non-locality is negligible at early times, then $ a\propto t^{\frac{1}{2}}$. This, in turn, implies $R = 0$, and the right hand side of the above equation vanishes. On the other hand, if the the scale factor evolution changes to a different power-law $a\propto t^s$ at a time $t_{\rm eq}$, then the above integration could be performed explicitly:
\begin{equation}\label{nonlocalDWs}
     \frac{1}{\Box}R(\bar{t}) = -\frac{6s\left(2s-1\right)}{\left(3s-1\right)}\left[\log{\Bar{t}} +\frac{1}{3s-1}\left(-1 + \Bar{t}^{3s-1}  \right) + \frac{1}{4}  \right]\;,
\end{equation}
where $\Bar{t}=t/t_{\rm eq}$. It is straightforward to realize from the above equation that the non-local contribution, which was vanishing during radiation domination, diverges at least logarithmically with time for a different power law ($s \neq 1/2$) and might induce an accelerated expansion. 

A different non-local correction to the Einstein-Hilbert Lagrangian can be obtained as the effective action of a non-minimally coupled scalar field. For example, non-local corrections emerge naturally within the framework of Sakharov's induced gravity  \cite{Sakharov:1967pk}, see also \cite{Zee:1980sj, Adler:1982ri, Visser:2002ew, Shapiro:2008sf}. In this theory one is able to have standard gravity, i.e. the Einstein Hilbert Lagrangian density, plus corrections induced by spontaneous symmetry breaking of the scalar field. We will see the details of this construction in Sec.~\ref{inducedgravity}, but let us anticipate that the resulting effective action is of the type:
\begin{equation}\label{effreefaction2}
    S_{\rm eff} = \frac{1}{2\kappa}\int d^4x\sqrt{-g}\left[\left(R - 2\Lambda\right) + Rf\left(\frac{1}{\Box + \beta}R\right)\right]\;,
\end{equation}
which recovers the action of the DW model for $\beta=0$. However, in the general case the non-local term is essentially the Green function for a Klein Gordon field with squared mass $\beta$ acting on the Ricci scalar.
As we are going to show, a non-vanishing $\beta$ can dramatically change the evolution of the non-local content. In particular, the picture suggested by Eqs.~\eqref{inverseboxR} and \eqref{effreefaction2} of a diverging field does not (necessarily) apply anymore, with interesting phenomenological consequences. 

The structure of the paper is the following: in Sec.~\ref{inducedgravity} we review Sakharov's construction, and show how it naturally leads to an effective action of the form \eqref{effaction2}, with $f(x) \propto x$. In Sec. \ref{localizedth} we derive a localized version of the action \eqref{effaction2} and the field equations for a general $f$. In Sec.~\ref{cosmology} we specialize our investigation to a flat FLRW background and discuss few interesting phenomenological implications, focusing on the main differences with respect to other non-local theories. Finally, in Sec.~\ref{conclusion} we summarize and elaborate our results.


\section{Sakharov Induced Gravity}\label{inducedgravity}

The goal of this section is to review Sakharov's induced gravity, and to show how it is effectively described by an action of the form \eqref{effaction2}. 
It is known that, in semi-classical gravity, in order to have a consistent (renormalizable and unitary) quantum field theory on curved space one needs to extend the usual Einstein-Hilbert action:
\begin{equation}\label{EHaction}
	S_{\rm EH} = \frac{1}{2\kappa}\int d^4x\sqrt{-g}(R - 2\Lambda)\;,
\end{equation}
where $\kappa \equiv 8\pi G$, to include higher-derivative contributions:
\begin{equation}\label{HDaction}
	S_{\rm HD} = \int d^4x\sqrt{-g}(\alpha_1C^2 + \alpha_2E_4 + \alpha_3 R^2 + \alpha_4\Box R)\;,
\end{equation}
where $C^2$ is the square of the Weyl tensor and $E_4$ is the Gauss-Bonnet term; $\alpha_{1,2,3,4}$ are dimensionless parameters. Moreover, Newton's constant $G$ and the cosmological constant $\Lambda$ in the Einstein-Hilbert action \eqref{EHaction} have to be intended as ``bare'' (as well as $\alpha_{1,2,3,4}$). In fact, they acquire radiative corrections due to the quantum nature of the matter fields. 
The two actions \eqref{EHaction} and \eqref{HDaction} form together the ``vacuum'' action, to which one has to add the matter one, in order to have the complete theory. See e.g. Ref.~\cite{Buchbinder:2021wzv} for a textbook review of these concepts.

The starting point of Sakharov's construction is to assume the absence of the vacuum action, and consider instead \textit{only} a simple matter model consisting of a complex scalar field $\varphi$ with action:
\begin{equation}\label{phiaction}
	S_{\varphi} = \int d^4x\sqrt{-g}\left[g^{\mu\nu}\partial_\mu\varphi^*\partial_\nu\varphi - V(\varphi,\varphi^*)\right]\;,
\end{equation}
where the Higgs-like potential $V(\varphi)$:
\begin{equation}\label{Vpot}
	V(\varphi,\varphi^*) = - \mu_0^2\varphi^*\varphi - \xi R\varphi^*\varphi + \lambda\left(\varphi^*\varphi\right)^2\;,
\end{equation}
is chosen in order to comply with renormalizability and in order to implement the spontaneous symmetry breaking mechanism, though modified by the presence of the $\xi R|\varphi|^2$ term. 
In other words, one assumes that gravity is no more a fundamental interaction but still retains the non-minimal coupling of the scalar field to curvature. Thanks to the latter, one is able to have gravity induced by spontaneous symmetry breaking. To illustrate this explicitly, let us consider the only equation of motion of the theory, obtained via variation with respect to $\varphi^*$ of the action \eqref{phiaction}:
\begin{equation}\label{eomvarphi}
    -\Box\varphi + \mu_0^2\varphi + \xi R\varphi - 2\lambda\varphi^3 = 0\;.
\end{equation}
In absence of minimal coupling, i.e. $\xi = 0$, one obtains the value:
\begin{equation}\label{flatsolution}
    \varphi_0^2 = \frac{\mu_0^2}{2\lambda}\;,
\end{equation}
and thus a continuum of classical vacua with $U(1)$ symmetry. However,  once that the scalar field relaxes around one of these vacuum state, the symmetry is broken and a Nambu–Goldstone boson appears. If the scalar field is coupled to a massless gauge vector field, after the symmetry breaking the latter acquires mass, realizing the well-known Higgs mechanism. 

On the other hand, if $\xi \neq 0$, there is no constant solution for the classical vacuum (unless $R$ is constant, too).
Let us assume:
\begin{equation}\label{approx}
	\xi R \ll \mu_0^2\,,\,\lambda\varphi^2\,.
\end{equation} 
Following Ref.\cite{Gorbar}, we can then treat $\xi R$ as a small perturbation and solve Eq.~\eqref{eomvarphi} by successive approximations. That is, assume an expansion:
\begin{equation}\label{phiexp}
    \varphi = \varphi_0 + \varphi_1 + \varphi_2 + \dots\;,
\end{equation}
where $\varphi_0$ is the flat space solution \eqref{flatsolution} and $\varphi_n = O[(\xi R)^n]$.  Keeping just the first order we obtain:
\begin{equation}\label{varphi1sol}
    \varphi_1 = \frac{\xi\varphi_0}{\Box + 4\lambda\varphi_0^2} R\;.
\end{equation}
Now, if we replace the above solution, up to second order, into the action \eqref{phiaction}, we obtain:
\begin{equation}\label{indaction}
    S_{\rm ind} = \int d^4x\sqrt{-g}\left[\lambda \varphi_0^4 + \xi R\varphi_0^2 + \xi^2\varphi_0^2R\frac{1}{\Box + 4\lambda \varphi_0^2}R\right]\;.
\end{equation}
This is an effective action, valid within the approximation made in Eq.~\eqref{approx}, in which gravity is induced (note the Einstein-Hilbert-like term $\xi R\varphi_0^2 $).\footnote{In other words, we found that when the scalar field oscillates around one of its classical vacua, its fluctuations in curved space-time produce an effective gravitational action. Notice that the notion of spontaneous symmetry breaking is well defined only for the constant, zero-th order solution $\phi_0$ in flat space. Thus, strictly speaking, in the present case is meaningful only at perturbative level.}

In particular, if gravity is totally induced, up to first order in $\xi$ we recover the standard Einstein-Hilbert action, with induced gravitational and cosmological constants [compare with Eq.~\eqref{EHaction}]:
\begin{equation}
    \xi \varphi_0^2 = \frac{1}{2\kappa_{\rm ind}}\;, \qquad -2\Lambda_{\rm ind} = \frac{\lambda \varphi_0^4}{\xi\varphi_0^2} = \frac{\lambda \varphi_0^2}{\xi}\;.
\end{equation}
Whilst the above action effectively describes  General Relativity plus non-local corrections, it is unfortunately inconsistent with cosmological observations. Since $\lambda$ must be positive, otherwise the spontaneous symmetry breaking mechanism could not be implemented, the constant contribution $\lambda \varphi_0^4$ is positive. In the Minimal Standard Model, the vacuum expectation value of the Higgs field is $\varphi_0 \simeq 246$ GeV and the measured Higgs mass is of order $\simeq 125$ GeV. So, $\lambda = \mathcal{O}(1)$. Then, we have a contribution to the cosmological constant density of order $\simeq 10^8$ GeV$^4$, against the observed value of $\simeq 10^{-47}$ GeV$^4$. Moreover, since $\xi$ has to be positive in order to have a positive Newton's constant, and thus gravity as an attractive force, we see that the induced cosmological constant has the wrong sign. We would have a huge, negative cosmological constant, totally incompatible with observation. Whereas the problem of the sign could be addressed by redefining the Lagrangian up to a minus sign, the problem of the huge value cannot be solved, if gravity is totally induced.   
Whilst this is a strong argument against the interpretation of gravity as a totally induced phenomenon, one could interpret Eq.~\eqref{indaction} as a non-local correction on top of a non-vanishing vacuum Einstein-Hilbert Lagrangian \eqref{EHaction} as follows:
\begin{equation}\label{effaction}
    S_{\rm eff} = \int d^4x\sqrt{-g}\left[\left(\frac{1}{2\kappa}R + \xi R\varphi_0^2\right) + \left(\lambda \varphi_0^4 - \frac{\Lambda}{\kappa}\right) +  \xi^2\varphi_0^2R\frac{1}{\Box + 4\lambda \varphi_0^2}R\right]\;.
\end{equation}
On the basis of the discussion above, since $\varphi_0 \simeq 246$ GeV and $1/\kappa $ is of the order of the Planck mass, we can neglect the correction induced on the gravitational constant.\footnote{This is possible provided that $\xi$ is of order one, or not too large, which must be the case by construction of the effective action.} For the cosmological constant,\textit{ in order for the above to be compatible with observation}, we need to assume that the bare value $\Lambda$ almost cancels the induced one, implying that this model is helpless against the  fine-tuning problem of the cosmological constant.\footnote{Exacerbating the fine-tuning, we may also think that $\Lambda = \kappa\lambda\varphi_0^4$, thereby making the effective cosmological constant vanishing. In this case, the non-local term only would drive the cosmic accelerated expansion.} 

Under these assumptions we can rewrite the effective action as follows:
\begin{equation}\label{effaction2}
    S_{\rm eff} = \int d^4x\sqrt{-g}\left[\frac{1}{2\kappa}\left(R - 2\Lambda\right) + \beta_1R\frac{1}{\Box + \beta_2}R\right]\;,
\end{equation}
where we have redefined the cosmological constant:
\begin{equation}
	\lambda \varphi_0^4 - \frac{\Lambda}{\kappa} \longrightarrow - \frac{\Lambda}{\kappa}\,,
\end{equation}
in order to avoid carrying on fastidious subscripts, and with:
\begin{equation}
	\beta_1 \equiv \xi^2\varphi_0^2\;, \qquad \beta_2 \equiv 4\lambda \varphi_0^2\,,
\end{equation}
both of order $10^4$ GeV$^2$, but not necessarily equal.\footnote{Note that the existence of other Higgs bosons related to spontaneous symmetry breaking taking place at higher energies is not ruled out. Therefore, $\beta_1$ and $\beta_2$ can be also thought of being larger.} 

As promised, we found (following Refs.~\cite{Buchbinder:2021wzv,Gorbar}) a new kind of non-local term featuring an unusual $\beta_2$ contribution, which is a particular case of the action \eqref{effaction2} with the simplest non-trivial choice for the free function $f(x) = \beta_1 x$, and with $\beta = \beta_2$.


\section{The localized theory}\label{localizedth}

When cosmological implications of a non-local theory are addressed, it is often helpful to work within a localized formulation of the theory, following the strategy adopted in Ref. \cite{Nojiri:2007uq} for the DW model \cite{Deser:2007jk}. 

The localization procedure goes as follows. Let us introduce two auxiliary fields $U$ and $V$ such that: \begin{equation}\label{indactionloc}
    S = \frac{1}{2\kappa}\int d^4x\sqrt{-g}\left(R - 2\Lambda\right) + \frac{1}{2\kappa}\int d^4x\sqrt{-g}\left[Rf + V\left(\Box U + \beta U - R\right)\right] + S_m\;.
\end{equation}
The role of $V$ is that of a Lagrange multiplier, and the variation of the action with respect to it gives us the equation of motion of the localized field $U$:
\begin{equation}\label{Ueq}
    \Box U + \beta U = R\;.
\end{equation}
The variation of the action with respect to $U$ gives us the equation of motion for $V$:
\begin{equation}\label{Veq}
    \Box V + \beta V = -\Bar{f}R\;,
\end{equation}
where $\Bar{f} \equiv \partial f/\partial U $. Once again, the extra $\beta$ term in the equations for the localized fields is a new feature of this class of non-local gravity models and we will pay special attention to it in the following.\footnote{This extra term might call to mind how Einstein introduced the cosmological constant in his seminal work \cite{Einstein:1917ce}, as a correction to the Poisson equation.} 
Finally, variation with respect to the metric yields the modified Einstein equations:\footnote{Note that, being in the action a non-minimal coupling of gravity with the localized field $U$ one might also adopt in the present calculation the Palatini formalism. We pursue this path in another paper.} 
\begin{equation}\label{efe}
\begin{split}
G_{\mu\nu}\left(1 + f -V\right) +g_{\mu\nu}\Lambda +\frac{g_{\mu\nu}}{2}\left(\partial_\rho U \partial^{\rho}V\right) -\frac{\beta}{2}g_{\mu\nu}UV +\mathcal D_{\mu\nu}\left(f-V\right) -\frac{1}{2}\left(\partial_\mu U\partial_\nu V + \partial_\mu V\partial_\nu U \right)=\kappa T_{\mu\nu}\,,
\end{split}
\end{equation}
where $G_{\mu\nu}$ is the Einstein tensor and we have defined the operator:
\begin{equation}
	\mathcal D_{\mu\nu} \equiv g_{\mu\nu}\Box - \nabla_\mu\nabla_\nu\,,
\end{equation}
for simplicity of notation. The energy-momentum tensor of matter is, as usual, defined as:
\begin{equation}
	T_{\mu\nu} \equiv -\frac{2}{\sqrt{-g}}\frac{\delta S_m}{\delta g^{\mu\nu}}\,.
\end{equation}
Note that our effective action \eqref{effaction} has only two propagating degrees of freedom, those of the metric. Now, after localization, it seems that we have gained two extra scalar degrees of freedom, $U$ and $V$. This is strange, especially if we consider that $V$ is a mere Lagrange multiplier. As extensively discussed in Ref.~\cite{Belgacem:2017cqo}, $U$ and $V$ might give rise to spurious degrees of freedom if we take into account the general solutions to Eqs.~\eqref{Ueq} and \eqref{Veq}, because these contain the homogeneous solutions:
\begin{equation}
    \Box U_{\rm hom} + \beta U_{\rm hom} = 0\;, \qquad \Box V_{\rm hom} + \beta V_{\rm hom} = 0\;.
\end{equation}
However, these homogeneous solutions are not required for passing from the action \eqref{indaction} to \eqref{indactionloc}.\footnote{It would be like adding a zero to the Lagrangian density in the action \eqref{indaction} and then demanding the existence of a scalar field such that $\Box U_{\rm hom} + \beta_2 U_{\rm hom} = 0$. This seems a pretty nonsensical thing to do.} Therefore, we need \textit{particular solutions} to Eqs.~\eqref{Ueq} and \eqref{Veq}, for fixed boundary conditions.\footnote{In other words, specifying the boundary conditions allow us to consider only the retarded Green function of the non-local differential operator, which makes the emerging ghost-like behaviour of the localized field harmless. However, it has been argued in Ref.\cite{Zhang:2016ykx} that a mixing of the advanced and retarded Green functions is unavoidable at the level of the equation of motion and the non-local theories might intrinsically suffer of a acausality problem. The phenomenology arising from a negative $\beta$ might resolve the latter problem since the auxiliary field $U$, as we are going to show, its not necessarily classically unstable in this scenario.} These boundary conditions might be considered as our ``degrees of freedom'', but not propagating ones. We stress that the above arguments are heuristic, and the proper way to count the number of degrees of freedom of the localized theory is through a Hamiltonian analysis.

Note also that the induced corrections are emerging from the fundamental action in Eq.~\eqref{phiaction}, so that we do have a scalar degree of freedom beyond the metric ones. Has this scalar degree of freedom been lost? In some sense yes, because the non-local action \eqref{indaction} is effective, i.e. it has been obtained by fixing $\varphi$ to its vacuum expectation value (corrected by $\xi$), so that it is no more a degree of freedom.


\section{Induced cosmology}\label{cosmology}
\subsection{Background equations}
In this section we investigate the effect of the induced non-local term of action \eqref{indaction} to cosmology. To this purpose, we adopt a spatially flat FLRW metric
\begin{equation}\label{flatFLRW}
    ds^2 = -dt^2 + a(t)^2\delta_{ij}dx^idx^j\;,
\end{equation}
and assume a perfect fluid energy momentum tensor
\begin{equation}
	T_{\mu\nu} = \sum_i(\rho_i + p_i)u_\mu u_\nu + p_ig_{\mu\nu}\,,
\end{equation}
where $i = m,r,\Lambda$, i.e. a non-interacting mixture of  matter, radiation, and a cosmological constant.
Using the e-fold number $N=\log{a}$ as time parameter the metric field equations \eqref{efe} can be written as follows:
\begin{equation}\label{feqe}
\left(1+f-V\right) = \frac{\Omega_m + \Omega_r +\Omega_\Lambda}{h^2}  +\frac{1}{6}\left(U'V' - \frac{\beta}{h^2}UV \right)-\left(f'-V'\right)\;,
\end{equation}
\begin{equation}\label{acceq}
    -\left(2\alpha +3\right)\left(1+f-V\right)=3\left(\frac{w\Omega_m + w\Omega_r -\Omega_{\Lambda}}{h^2}\right) +\frac{1}{2}\left(U'V' +\frac{\beta}{h^2}UV \right) + \left[f'' -V'' +f'\left(\alpha +3\right) -V'\left(\alpha +3\right)\right]\;,
\end{equation}
where we have introduced the normalized Hubble factor $h \equiv H/H_0$ and its 
 logarithmic derivative $\alpha \equiv h'/h$, the normalized energy densities $\Omega_i \equiv 8\pi G \rho_i /3H_0^2$, the  and redefined the parameter $\beta\rightarrow \beta/H_0$. Using these variables the equations of motion for the non-local fields become:
\begin{equation}\label{Ueqflrw}
    U'' + \left(3+\alpha\right)U' - \frac{\beta}{h^2} U = -6\left(2+\alpha\right)\;,
\end{equation}
\begin{equation}
    V'' + \left(3+\alpha\right)V' - \frac{\beta}{h^2}V= +6\Bar{f}\left(2+\alpha\right)\;.
\end{equation}

\subsection{Qualitative Behaviour of $U$ across the cosmological evolution}
\subsubsection*{Initial conditions: Radiation domination}
Let us stress once more that in order to avoid the propagation of spurious degrees of freedom, we should discard the homogeneous part of the above field equations and consider only particular,  retarded solutions. In order to do so, we have to specify initial conditions for the fields at the initial time $N=N_i$. If at the initial time $N_i$ we also want the universe to be compatible with the $\Lambda$CDM evolution, i.e. $h^2(N_i) = \Omega_m (N_i) + \Omega_r (N_i) +\Omega_\Lambda$,  then the Friedmann-like equation \eqref{feqe} gives:
    \begin{equation}
   f(N_i)-V(N_i)= \frac{1}{6}\left[U'(N_i)V'(N_i)-\frac{\beta}{h^2(N_i)}U(N_i)V(N_i)\right] - f'(N_i) + V'(N_i)\;.
\end{equation}
If we want to recover the standard Friedmann equation at early times, we can satisfy the above constraint with the choice $U(N_i)=U'(N_i)=V'(N_i)=V(N_i) = 0$ and assuming that $f$ is a homogeneous function of $U$. This choice is also compatible with the Klein Gordon equation \eqref{Ueqflrw} since during radiation domination we have $\alpha = -2$. Therefore, as it is common for many similar models, non-localities become important only at a later stage of the evolution of the universe. This is desirable (and not \textit{ad hoc}) if we do not want to spoil the successes of the standard model at early-times and, at the same time, justify why the expansion of the universe speeds up at late times.

\subsubsection*{Qualitative analysis: Matter domination}
\begin{figure}[h]
\includegraphics[scale=0.98,trim=0mm 0mm 0mm 0mm, clip]{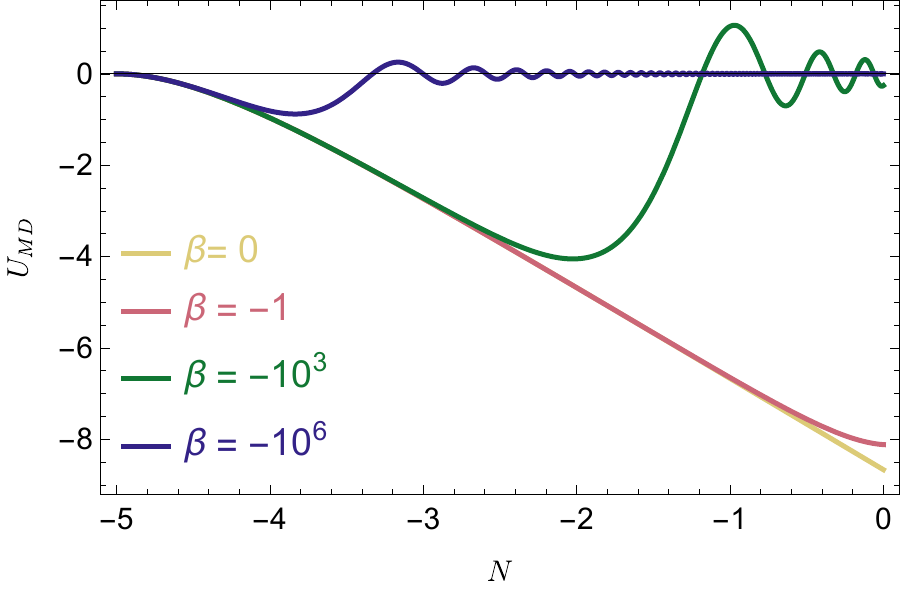}\;\includegraphics[scale=1,trim=0mm 0mm 0mm 0mm, clip]{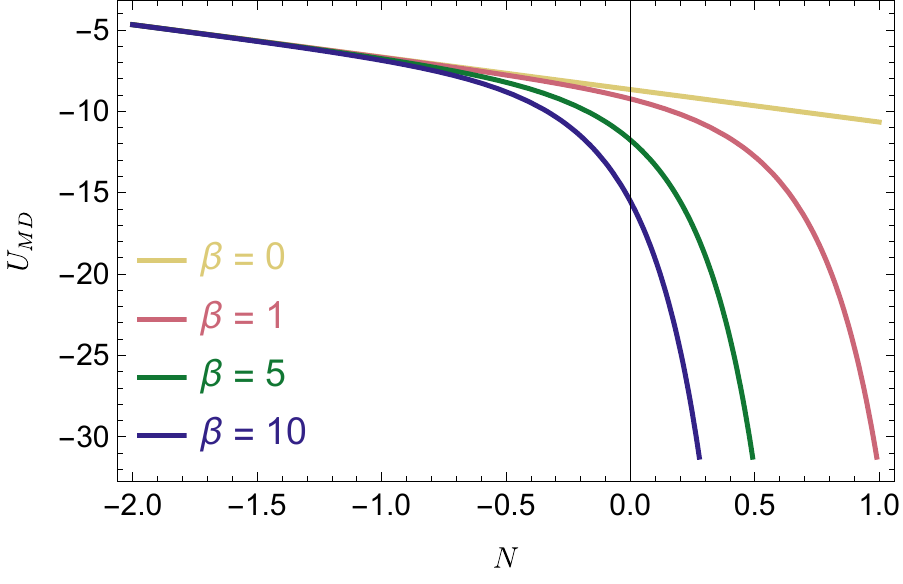}
	\caption{The evolution of the field $U$ in a matter-dominated background for different values of $\beta$ obtained from the analytical solutions of Eq.~\eqref{UeqMD}. Vanishing initial conditions for $U$ and $U'$ have been imposed at $N_i = -5$. \\
 \textit{Left:} A negative  $\beta < 0$ induces an oscillatory evolution for the field $U$. The largest the value of $|\beta|$, the quickest the field ceases its linear growth and starts to oscillate.\\
 \textit{Right:} a positive $\beta\geq 0 $ induces an exponential evolution for the field $U$, exacerbating its classical instability. The largest the value of $|\beta|$, the quickest the field transition from the linear to the exponential growth.}
	\label{Umb}
\end{figure}

When $\Omega_m$ is not anymore negligible with respect to $\Omega_r$ radiation domination ends, $\alpha \neq 0$ and the non-local field $U$ starts to evolve. If the latter grows slowly enough for the Universe to become dominated by its pressure-less matter content (i.e. dust dominates also over the non-localities), then $h^2 \propto e^{-3N}$ and $\alpha = -3/2$. In this regime the  equation for the field $U$ becomes
\begin{equation}\label{UeqMD}
    U'' +\frac{3}{2}U' - \beta e^{3N}U = -3\;,
\end{equation}
which admits analytic solutions. For $\beta=0$ the solution is $U\approx 2N+ k$, where $k$ is a constant depending on the time at which vanishing initial conditions have been set. Thus, in the standard DW model and similar nonlocal theories, the field $U$ grows linearly during matter domination. However, a non-vanishing $\beta$ can severely change this picture. If $\beta \leq 0$, it induces oscillations in the field $U$ after it reaches a maximum value, as showed in Fig. \ref{Umb}. In particular, it is possible to choose $\beta$ in such a way that the field starts to oscillate arbitrarily close to the initial time $N_i$, thus suppressing the linear growth of $U$ before the end of matter domination. This feature is unique of this particular class of non-local models, and results in a number of interesting phenomenological implications, which we discuss in detail in Sec.~\ref{conclusion}. On the other hand, if $\beta \geq 0$, the field $U$ at some point starts to grow exponentially.  
\subsubsection*{Qualitative analysis: Late time evolution}
\begin{figure}[h]
\includegraphics[scale=0.95,trim=0mm 0mm 0mm 0mm, clip]{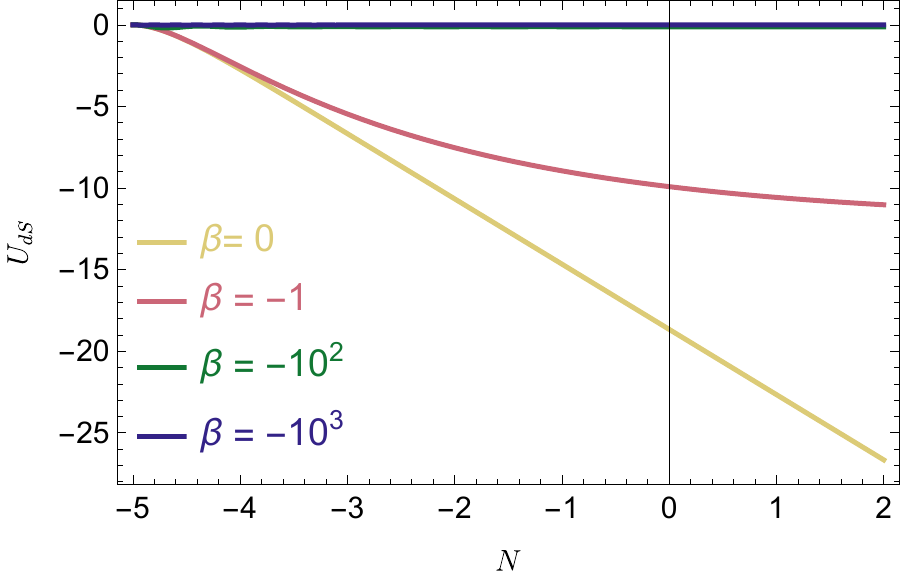} \; \includegraphics[scale=0.95,trim=0mm 0mm 0mm 0mm, clip]{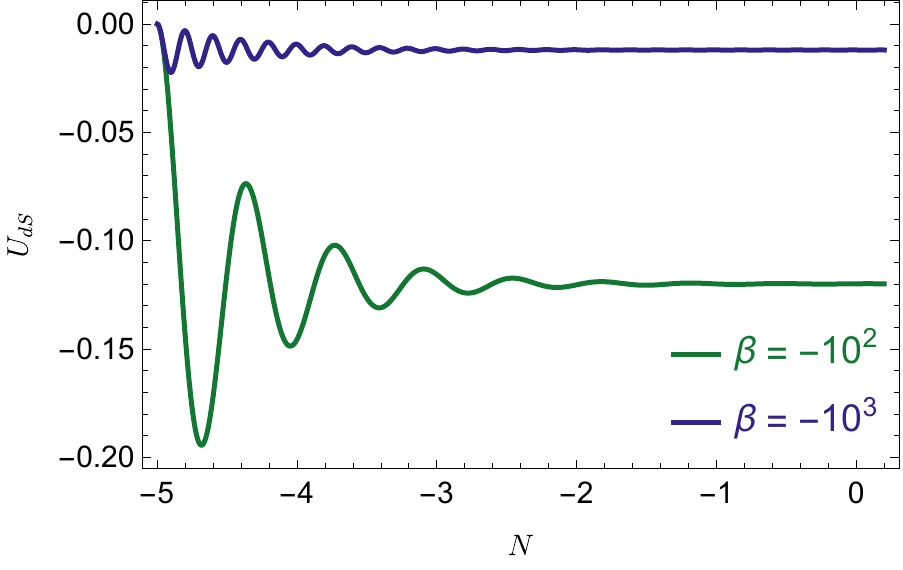}\;
	\caption{\textit{Left}: The evolution of the field $U$ in a de Sitter background for different values of $\beta \leq 0$ predicted by the analytical solutions of Eq.~\eqref{UeqdS}. Vanishing initial conditions for $U$ and $U'$ have been imposed at $N_i = -5$. It is straightforward to realize that the presence of a negative $\beta$ stops the linear growth of the field $U$, which relaxes asymptotically towards a constant value. The largest the value of $|\beta|$ is, the quickest the field relaxes to a constant value.\\
 \textit{Right}: A zoom on the evolution of $U$ for $\beta=-10^{2}$ and $\beta=-10^{3}$, showing that $U$ evolves through decaying oscillations towards a constant value.}
	\label{UDsnb}
\end{figure}

\begin{figure}[h]
\includegraphics[scale=0.95,trim=0mm 0mm 0mm 0mm, clip]{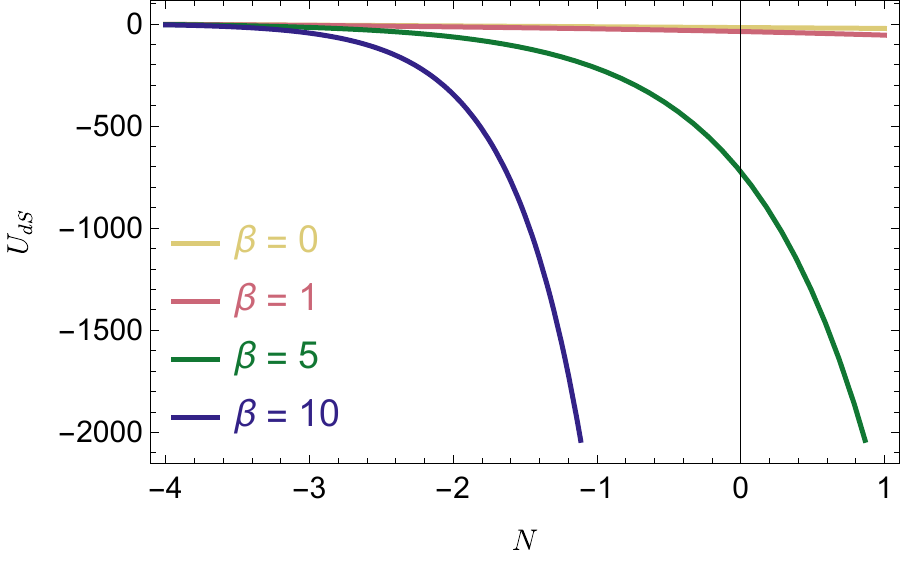} \; \includegraphics[scale=0.95,trim=0mm 0mm 0mm 0mm, clip]{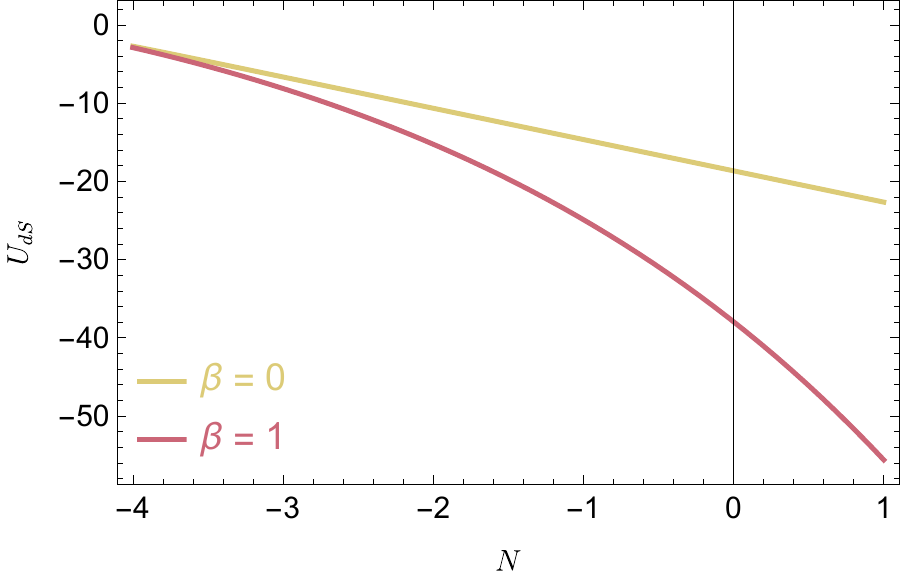}\;
	\caption{\textit{Left}: The evolution of the field $U$ in a de Sitter background for different values of $\beta \geq 0$ predicted by the analytical solutions of Eq.~\eqref{UeqdS}. Vanishing initial conditions for $U$ and $U'$ have been imposed at $N_i = -5$. It is straightforward to realize that the presence of a negative $\beta$ induces an exponential growth for the field $U$. The largest the value of $|\beta|$, the quickest the transition from the linear to the exponential evolution.\\
 \textit{Right}: A zoom on the evolution of $U$ for $\beta=0$ and $\beta=1$, showing that for smaller values of $|\beta|$ the transition to the exponential regime occurs at later times.}
	\label{UDspb}
\end{figure}

Because of the divergence of the field $U$, the late time evolution of the universe for most non-local gravity models is usually different from the $\Lambda$CDM one. In particular, if $U$ enters explicitly the field equations, it has been shown in Ref.~\cite{Giani:2019xjf} that at late-times the non-local contributions behave as a phantom fluid, and the asymptotic effective equation of state parameter approaches asymptotically from below the one of a cosmological constant $w = -1$. On the other hand, including the $\beta$ term could suppress the growth of $U$, as we have shown earlier for $\beta < 0$ during matter domination. Let us suppose that the non-local energy density contribution in Eq. \eqref{feqe} is still negligible compared to the cosmological constant one $\Omega_\Lambda$, and that the latter dominates over the dust one. This regime is thus effectively described by a de Sitter background, for which $h \simeq$ constant and $\alpha=0$. The Klein Gordon equation for $U$ in this regime is given by:
\begin{equation}\label{UeqdS}
 U'' +3U' -\tilde{\beta} U = -12\;,   
\end{equation}
where we have defined $\Tilde{\beta}=\beta/h^2$. Some solutions for varying $\tilde{\beta}$ are plotted in Figs.~\ref{UDsnb} and \ref{UDspb} for negative and positive values of $\beta$. In the former case the field $U$ approaches asymptotically a constant value, which is also a novel feature of this class of non-local models, in net contrast with the approximately linear growth that $U$ experiences on a de Sitter background when $\beta=0$. Similarly to the matter dominated case, we notice that a positive $\beta$ induces an exponential growth of the field $U$. 


\subsection{Simplest non-trivial model: $f(U) = \beta_1 U$}

Let us now specialize to the action functional \eqref{effaction2}, i.e. let us choose $f(U) = \beta_1 U$. It is straightforward to realize that in this case the Klein-Gordon equations \eqref{Ueq} and\eqref{Veq}, choosing vanishing initial conditions for both fields and scaling the Lagrange multiplier $V \rightarrow k\beta_1 V$ become identical, and therefore $V = -\beta_1 U$. The dynamical field equations hence can be written:
\begin{equation}\label{sakhamodfe}
    1+2\beta_1U = \frac{\Omega_m + \Omega_r +\Omega_\Lambda}{h^2} -\frac{\beta_1}{6}\left(U'^2 - \frac{\beta_2}{h^2}U^2 \right) -2\beta_1 U'\;, 
\end{equation}
\begin{equation}\label{sakhamodacceq}
    -\left(2\alpha +3\right)\left(1+2\beta_1 U\right)=3\left(\frac{w\Omega_m + w\Omega_r -\Omega_{\Lambda}}{h^2}\right) -\frac{\beta_1}{2}\left(U'^2 +\frac{\beta_2}{h^2}U^2 \right) + 2\beta_1\left[U'' +U'\left(\alpha +3\right)\right]\;,
\end{equation}
\begin{equation}\label{sakhamodkg}
    U'' +\left(3+\alpha\right)U' -\frac{\beta_2}{h^2} U =-6\left(2+\alpha\right)\;.
\end{equation}

\begin{figure}[h]

\includegraphics[scale=1,trim=0mm 0mm 0mm 0mm, clip]{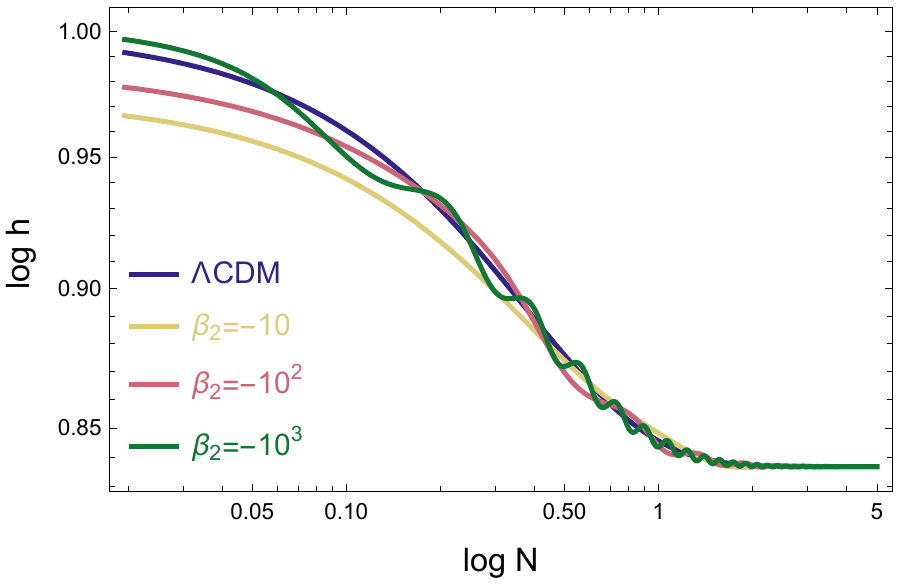}\;\includegraphics[scale=1,trim=0mm 0mm 0mm 0mm, clip]{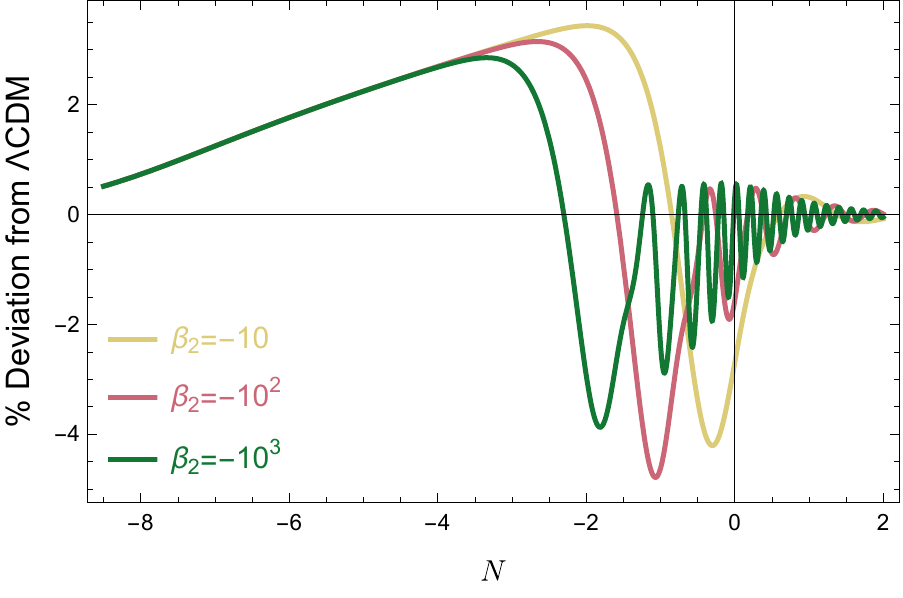}\;
	\caption{\textit{Left}: The evolution of $h$ obtained integrating numerically Eqs.~\eqref{sakhamodacceq}, and \eqref{sakhamodkg} for different values of $\beta_2\leq 0$. We assumed vanishing initial conditions for $U$ and $U'$ during radiation domination, at $N_i = -15$, and the present day energy densities $\Omega_m(0) =0.3 -9.2 \times 10^{-5} $, $\Omega_r(0) =9.2 \times 10^{-5}$, and $\Omega_\Lambda (0) = 0.7$. It is straightforward to realize that the presence of a negative $\beta_2$ induces an oscillatory behaviour for the Hubble factor, which however approaches asymptotically the same value as in the $\Lambda$CDM model. \\
    \textit{Right}: The relative difference between the $h$ predicted from the models with $\beta_2<0$ and the $\Lambda$CDM expansion history in percent units.}
	\label{hevologn}
\end{figure}
\begin{figure}[h]
\includegraphics[scale=1,trim=0mm 0mm 0mm 0mm, clip]{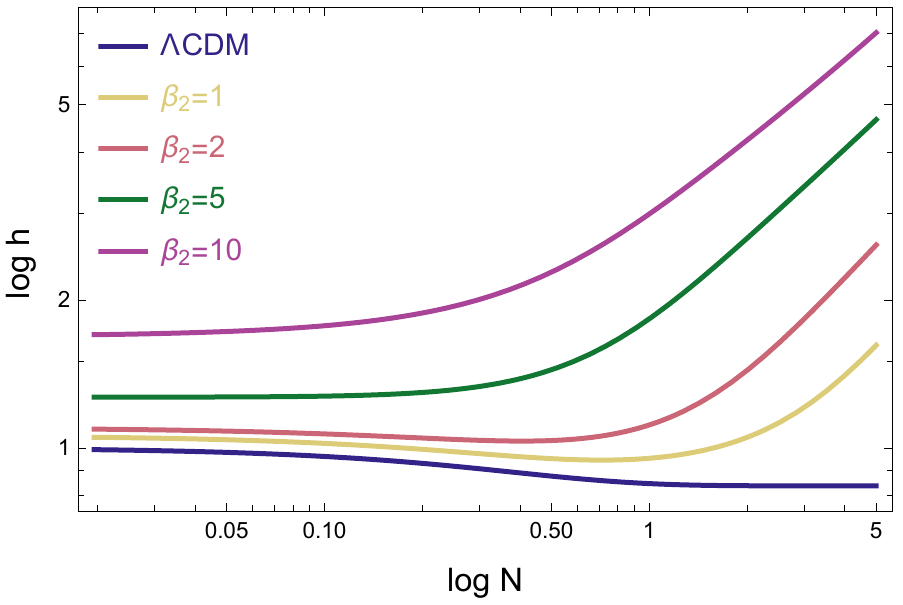}\;\includegraphics[scale=1,trim=0mm 0mm 0mm 0mm, clip]{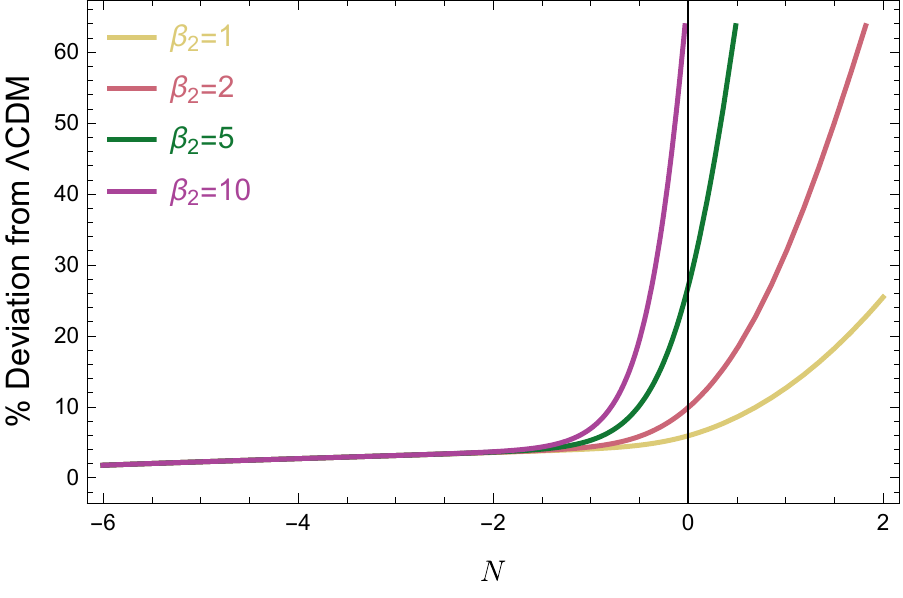}\;
	\caption{\textit{Left}: The evolution of $h$ obtained integrating numerically Eqs.~\eqref{sakhamodacceq} and \eqref{sakhamodkg} for different values of $\beta_2\geq0$. We assumed vanishing initial conditions for $U$ and $U'$ during radiation domination, at $N_i = -15$, and the present day energy densities $\Omega_m(0) =0.3 -9.2 \times 10^{-5} $, $\Omega_r(0) =9.2 \times 10^{-5}$, and $\Omega_\Lambda (0) = 0.7$. We see that the presence of a positive $\beta_2$ induces an exponential growth for $h$, with the transition to the exponential regime being faster for larger values of $|\beta_2|$.  \\
    \textit{Right}: The relative difference between the $h$ predicted from the models with $\beta_2\geq 0$ and the $\Lambda$CDM expansion history in percent units.}
	\label{hevologp}
\end{figure}

In Figs.~\ref{hevologn} and \ref{hevologp} we plot the evolution of the Hubble parameter in this model for different values of $\beta_2$ and a fixed value for $\beta_1 = 10^{-3}$. For negative $\beta_2 < 0$, the oscillatory behaviour of the non-local terms makes the model different but qualitatively similar to the $\Lambda$CDM behaviour at all times. On the other hand, we see that the effect of a positive $\beta_2$ is to exacerbate the classical instability of the non-local energy density contribution at late times.


\section{Discussion}\label{conclusion}

In this work we studied the cosmological impact of a non-local modification of the Einstein Hilbert action involving the operator $\left(\Box +\beta\right)^{-1}$ acting on the Ricci scalar. The presence of the $\beta$ term is a novelty in the landscape of proposed non-local modifications of gravity, and results in an interesting and rich phenomenology. In Sec.~\ref{localizedth} we derived the field equations for the theory described by the action \eqref{effreefaction2}, containing an arbitrary function $f$ of the aforementioned operator. On a fundamental level such an action can arise from a non-minimal coupling of the (quantum) matter Lagrangian with the space-time curvature. In particular, we shown in Sec.~\ref{inducedgravity} that a simple model with $f(x)\propto x$ is straightforwardly obtained in the contest of Sakharov induced gravity, see Eq.~\eqref{indaction}, from a complex scalar field with Higgs-like potential.
It is often convenient to work within a localized version of these non-local theories by defining the auxiliary field $U=\left(\Box +\beta\right)^{-1} R$, and a Lagrange multiplier $V$ which enforces the definition of $U$ at the level of the action. In terms of $U$ and $V$ we can rewrite the action in the form of a multi scalar-tensor theory, even if care must be taken in the choice of initial conditions to avoid the introduction of unphysical propagating degrees of freedom. 

From the phenomenological point of view we can distinguish between two classes of models depending on the sign of $\beta$. It is well known that for $\beta = 0$ the field $U$ evolves linearly with time (measured in units of e-folds $N$) during the matter dominated epoch, and therefore will eventually dominate over the matter content itself. For $\beta>0$ the behaviour is initially qualitatively similar, but at some point the field $U$ grows faster than linearly, thus exacerbating the usual classical instability typical of non-local theories.  
On the other hand, a negative $\beta$ results in a drastically different behaviour. Indeed, we shown in Sec.~\ref{cosmology} that for $\beta <0$ the field $U$ acquires an oscillatory behaviour and approaches at late times a constant value. These oscillations  start already in the matter dominated epoch, and thus may have important consequences on the formation and growth of large scales structures. For example, from Eq.~\eqref{sakhamodfe} we see that this model predicts an effective gravitational coupling $G_{\rm eff} = G_N/(1+2\beta_1 U)$. Since $U$ can oscillate between positive and negative values, the resulting picture is one where the universe, through its evolution, experiences alternating epochs of weaker and stronger gravity. 

We argue that such a rich phenomenology is particularly suited to address the cosmic tensions on the cosmological parameters $\sigma_8$ and $H_0$ \cite{DiValentino:2020vvd, Kazantzidis:2019nuh, Bernal:2016gxb,Abdalla:2022yfr,Verde:2019ivm}. Indeed, we proved that in this model we can have at the same time a faster (potentially phantom-like) expansion rate today (the oscillations on $U$ also imply an oscillating Hubble factor, see Fig.~\ref{hevologn}) and a weaker gravity regime at the epoch of structure formation. These two properties are, according to Ref.~\cite{Heisenberg:2022gqk}, necessary conditions to be satisfied in order to address the $H_0$ tension through late-times modifications of the expansion history without worsening the $\sigma_8$ tension. 
As argued in Ref.~\cite{Marra:2021fvf}, a time-varying gravitational coupling might also result in a lower luminosity of local supernovae, thus alleviating the Hubble tension. 
Furthermore, in Ref.~\cite{Lee:2022cyh} it has been shown that late-time resolutions of the $H_0$ tension require dynamical DE crossing the phantom line $w< -1$ which also need to have integrated energy density smaller than the one of the cosmological constant in $\Lambda$CDM. This last requirement could be potentially satisfied if the gravitational coupling varies with time. With a number of analysis \cite{Efstathiou:2021ocp,Vagnozzi:2019ezj,Camarena:2021jlr,Benevento:2020fev} indicating that late times phantomic DE alone is not enough to solve the Hubble tension, we argue that the phenomenology presented in this work makes this class of models particularly interesting.  
Despite the fact that it might be premature to interpret the $\sigma_8$ tension as a hint towards new physics, see for example Refs.~\cite{Efstathiou:2017rgv,Nunes:2021ipq}, recent analysis highlighted a discrepancy between the measured growth rate of structures and the one predicted by $\Lambda$CDM, see for example Ref.~\cite{Nguyen:2023fip} and Fig.11 of Ref.~\cite{Boubel:2023mfe}. In light of these results, we believe that a model predicting an oscillating gravitational coupling, as long as the average of the fluctuations evolves significantly between different redshift bins, might be particularly interesting.
Furthermore, whilst many non-local modifications of gravity are ruled out by constraints on the variation of $G$ from Lunar Laser Ranging (LLR) experiments, see for example Ref.~\cite{Belgacem:2018wtb}, if $U$ approaches a constant value at late times instead of diverging the latter constraints might not apply. 

It must be stressed that a negative $\beta$ is not compatible with the effective action of Eq.~\eqref{indaction}, since for the latter we have $\beta = 4\lambda \phi_0^2$ which must be positive for the spontaneous symmetry breaking to occur. Nevertheless, it is encouraging that this type of non-local terms can arise in such a simple and natural way, and one could conceive that a negative $\beta$ might be obtained using different potentials for the scalar field, or a different non-minimal coupling of the matter Lagrangian.  Our main take away is that the phenomenology described here provides an effective tool to modify the cosmological evolution only for a brief window of time somewhere between the matter and cosmological constant dominated epoch. It also provides a proof of concept that non-local modifications are not necessarily unstable, and in particular that non-local effects might only be important during matter domination rather than later. In order to fully understand their impact on structure formation one of course needs to study the evolution of linear perturbations, which might differ significantly from their evolution in the $\Lambda$CDM model, even if their background is arbitrarily close. See, for example, Refs.~\cite{Park:2017zls,Park:2012cp,Dodelson:2013sma}  where the issue of structure formation has been studied for the DW model. We plan to address these compelling questions in future investigations.

\subsection*{Acknowledgments}
LG acknowledge support from the Australian Government through the Australian Research Council Laureate Fellowship grant FL180100168.

\appendix

\bibliographystyle{unsrturl}
\bibliography{induce.bib}

\end{document}